%% file: miccai04.tex
\documentclass{llncs}
\usepackage{makeidx} 
\usepackage{amsmath,epsfig,xspace}
\usepackage{theorem}
\usepackage{amssymb}
\graphicspath{{./figures/}}
\input Input/alphabet.tex
\input Input/abrmath.tex

\input Input/abrege.tex
\input Input/beginend.tex

\def\bm#1{\mbox{\bf #1}}

\def\d#1{\,\hbox{d}#1}
\def\bs{\bar{s}}
\def\us{\underline{s}}
\def\fmin{f_{\mbox{min}}}
\def\fmax{f_{\mbox{max}}}
\def\expf#1{\mbox{exp}\left\{#1\right\}}
\def\argmin#1#2{\mbox{arg}\min_{#1}\left\{#2\right\}}
\def\argmax#1#2{\mbox{arg}\max_{#1}\left\{#2\right\}}
\def\ufb{\underline{\fb}}
\def\ugb{\underline{\gb}}
\def\uthetab{\underline{\thetab}}
\def\uvarepsilonb{\underline{\varepsilonb}}

\def\lra{\longrightarrow}
\def\fh{\widehat{f}}
\def\fbh{\widehat{\fb}}
\def\Rbh{\widehat{\Rb}}
\def\Sigmabh{\widehat{\Sigmab}}

\def\rbh{\widehat{\rb}}
\def\qbh{\widehat{\qb}}
\def\xbh{\widehat{\xb}}
\def\tbh{\widehat{\tb}}
\def\thetabh{\widehat{\thetab}}

\def\disp{\displaystyle}
\def\vsm{\vspace*{-12pt}}
\def\hsm{\hspace*{-3em}}

\def\pmata#1#2{\left(\barr{c} #1 \\ #2 \earr\right)}
\def\pmatb#1#2#3#4{\left(\barr{cc} #1 & #2 \\ #3 & #4 \earr\right)}

\def\th{\widehat{t}}
\def\xh{\widehat{x}}
\def\lambdah{\widehat{\lambda}}
\def\muh{\widehat{\mu}}
\def\sigmah{\widehat{\sigma}}
\def\alphah{\widehat{\alpha}}
\def\betah{\widehat{\beta}}
\def\tbh{\widehat{\tb}}
\def\mubh{\widehat{\mub}}
\def\sigmabh{\widehat{\sigmab}}
\def\alphabh{\widehat{\alphab}}

\def\Po{\textbf{P1\xspace}}
\def\Pt{\textbf{P2\xspace}}
\def\Pth{\textbf{P3\xspace}}
\def\Pf{\textbf{P4\xspace}}

\begin{document}

\pagestyle{headings}  

%
%
\title{A hidden Markov Model for image fusion and their joint segmentation in medical image computing}
%
%
\author{Olivier FERON and Ali MOHAMMAD-DJAFARI}

\institute{Laboratoire des signaux et syst\`emes, \\
Sup\'elec, plateau de Moulon,\\
3, rue Joliot Curie, \\
91192 Gif-sur-Yvette \\
\email{\{feron,djafari\}@lss.supelec.fr},
}
\maketitle
\begin{abstract}
\noindent
In this work we propose a Bayesian framework for fully automated image fusion and their joint segmentation. More specifically, we consider the case where we have observed images of the same object through different image processes or through different spectral bands. The objective of this work is then  to propose a coherent approach to combine these data sets and obtain a segmented image which can be considered as the fusion result of these observations. \\
The proposed approach is based on a Hidden Markov Modeling (HMM) of the images with common segmentation, or equivalently, with common hidden classification label variables which are modeled by the Potts Markov Random Field. We propose an appropriate Markov Chain Monte Carlo (MCMC) algorithm to implement the method and show some simulation results and applications. \\

\noindent
{\bf key words :} \\
Data fusion, Segmentation, multispectral images, HMM, MCMC, Gibbs Algorithm. \\
\end{abstract}

\section{Introduction}

\noindent
Data fusion and multi-source information has become a very active area of research in many domains : industrial nondestructive testing and evaluation (\cite{Gautier95b}), industrial inspection (\cite{Bass00}), and medical imaging \cite{Matsopoulos,Johnston96,Wang03,Reddick97,Ahmed02}. For example in magnetic resonance (MR) image segmentation, one can use multispectral images to accentuate the differencies in physical characteristics of the anatomical tissues. In this case we must fuse information of these multispectral images to obtain segmentation.\\
The main problem is how to combine the information contents of different sets of data $g_i(r)$. Very often the data sets $g_i$, and corresponding images $f_i$, do not represent the same quantities. A general model for these problems can be the following :
\beq
g_i(r)=[H_if_i](r) + \varepsilon_i(r), \quad i=1,\dots,M
\eeq
\noindent
where $H_i$ are the functional operators of the measuring systems, or a registration operators if the observations have to be registered. We may note that estimating $f_i$ given each set of data $g_i$ is an inverse problem by itself. In this work we propose to reconstruct images $f_i$ and to construct a common segmentation at the same time. This fused segmentation will be presented by an image $\zb$. 
\begin{figure}[h]
\centering
\begin{tabular}{ccc}
	\includegraphics[width=22mm,height=22mm]{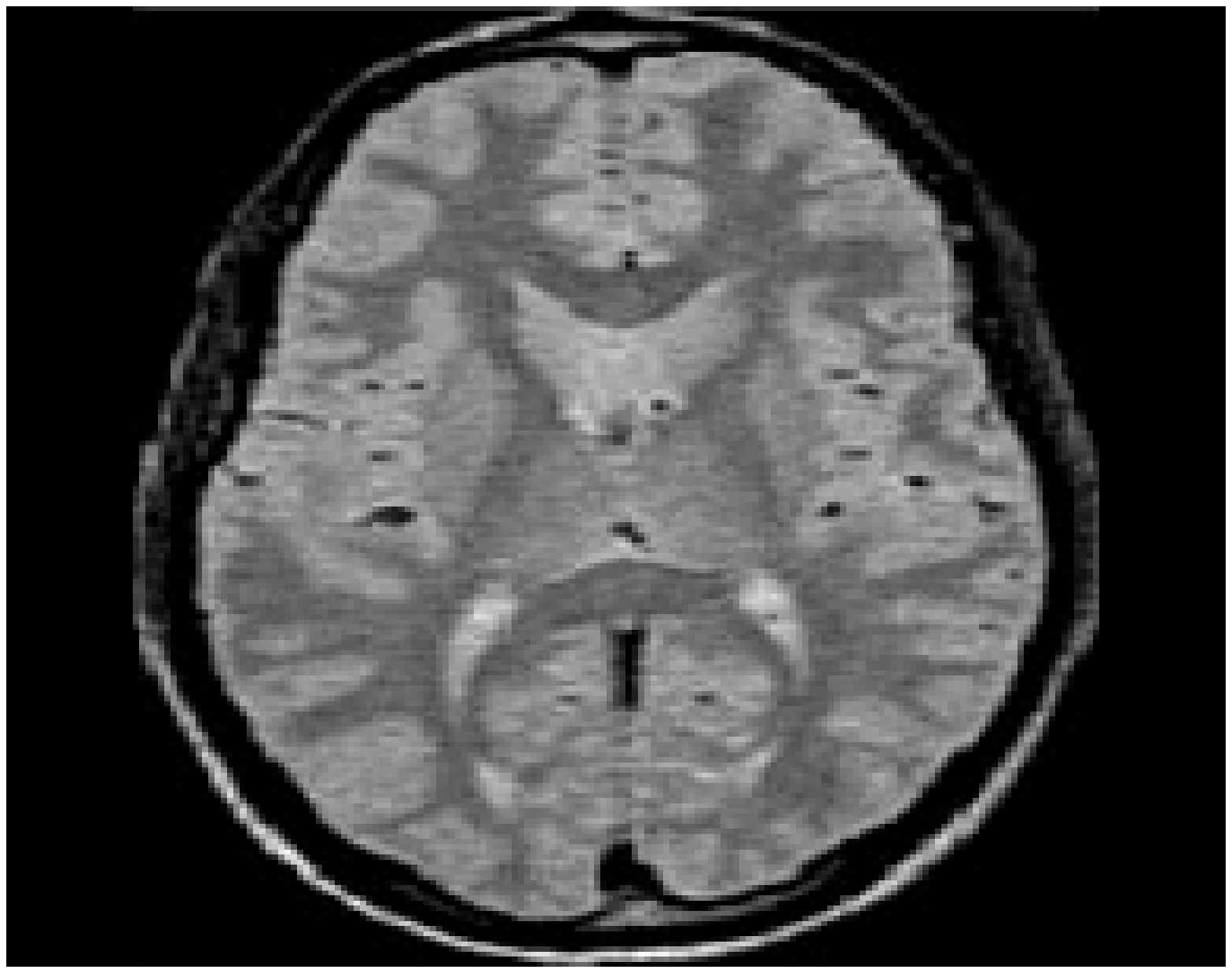} & 
	\includegraphics[width=22mm,height=22mm]{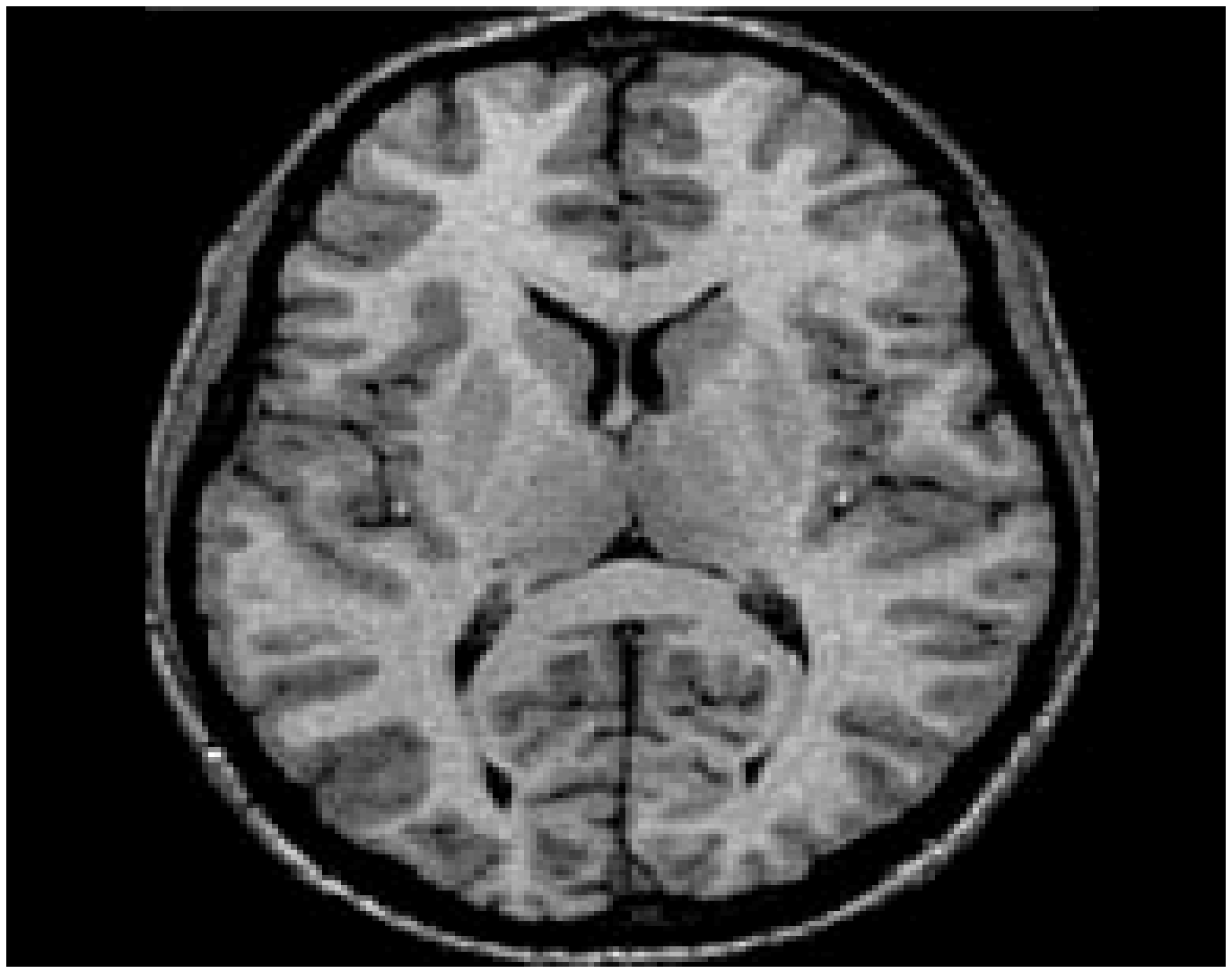} & 
	\includegraphics[width=22mm,height=22mm]{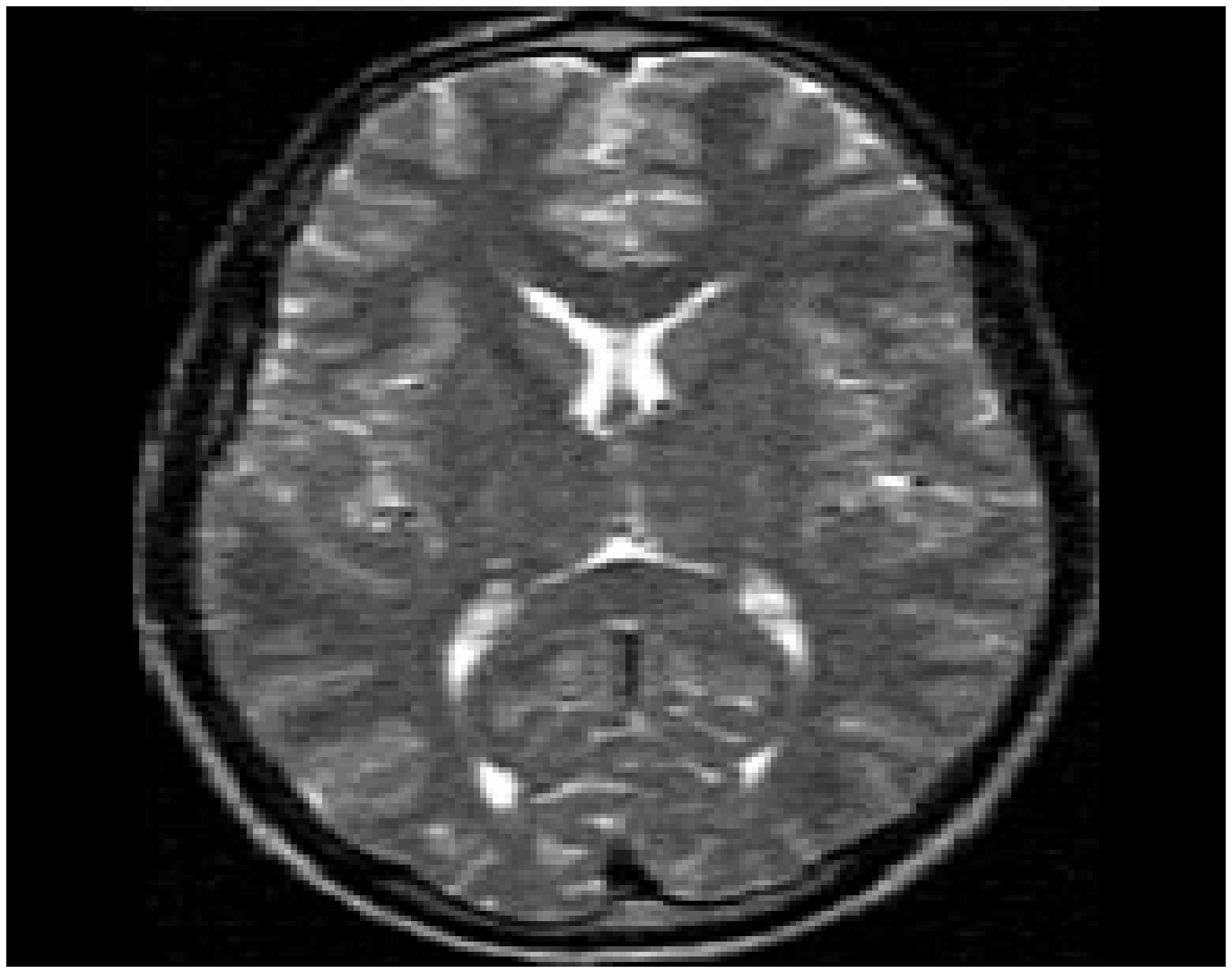} 
\\
 \multicolumn{3}{c}{(a)} 
\end{tabular}
\begin{tabular}{ccc}
	\includegraphics[width=22mm,height=22mm]{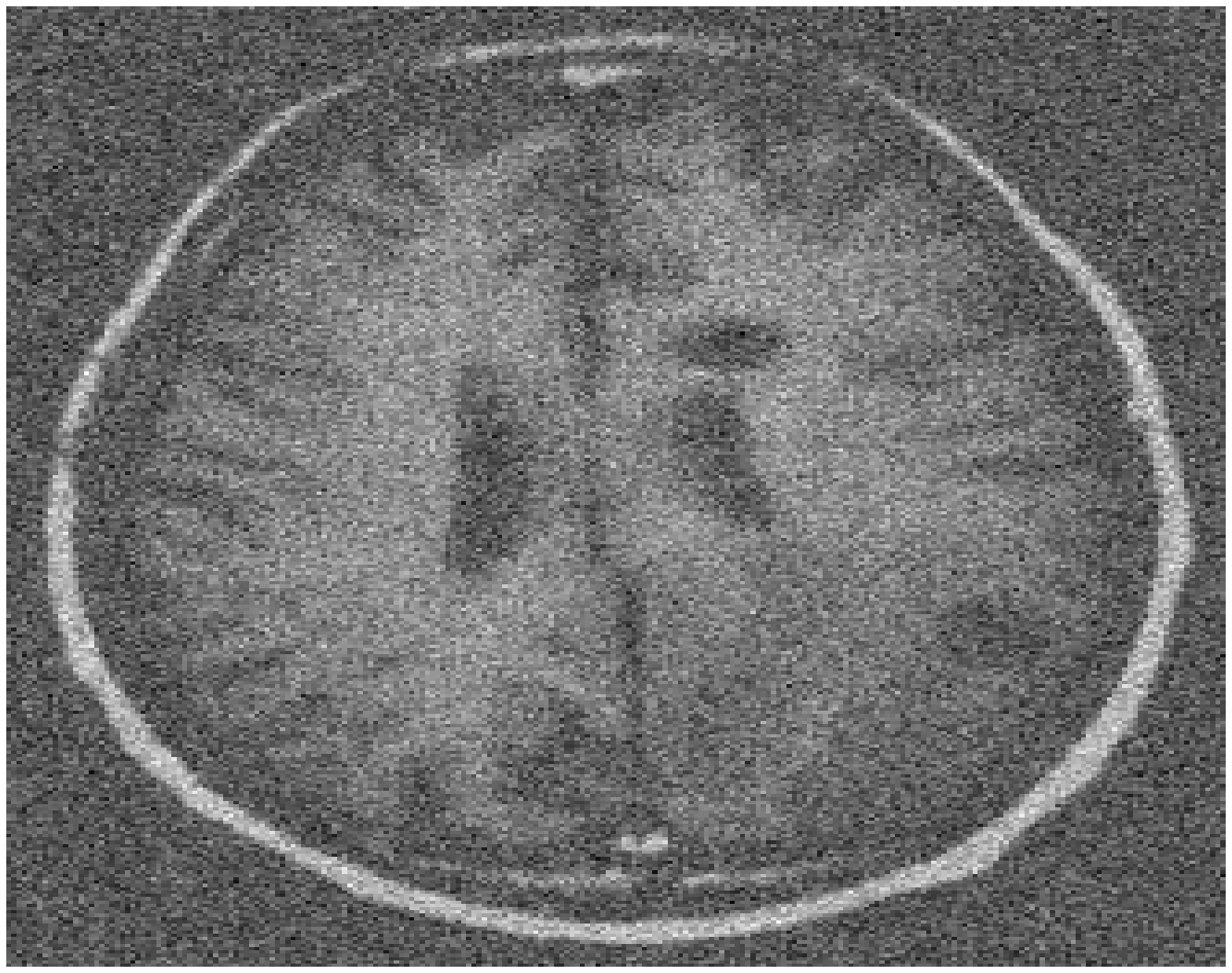} & 
	\includegraphics[width=22mm,height=22mm]{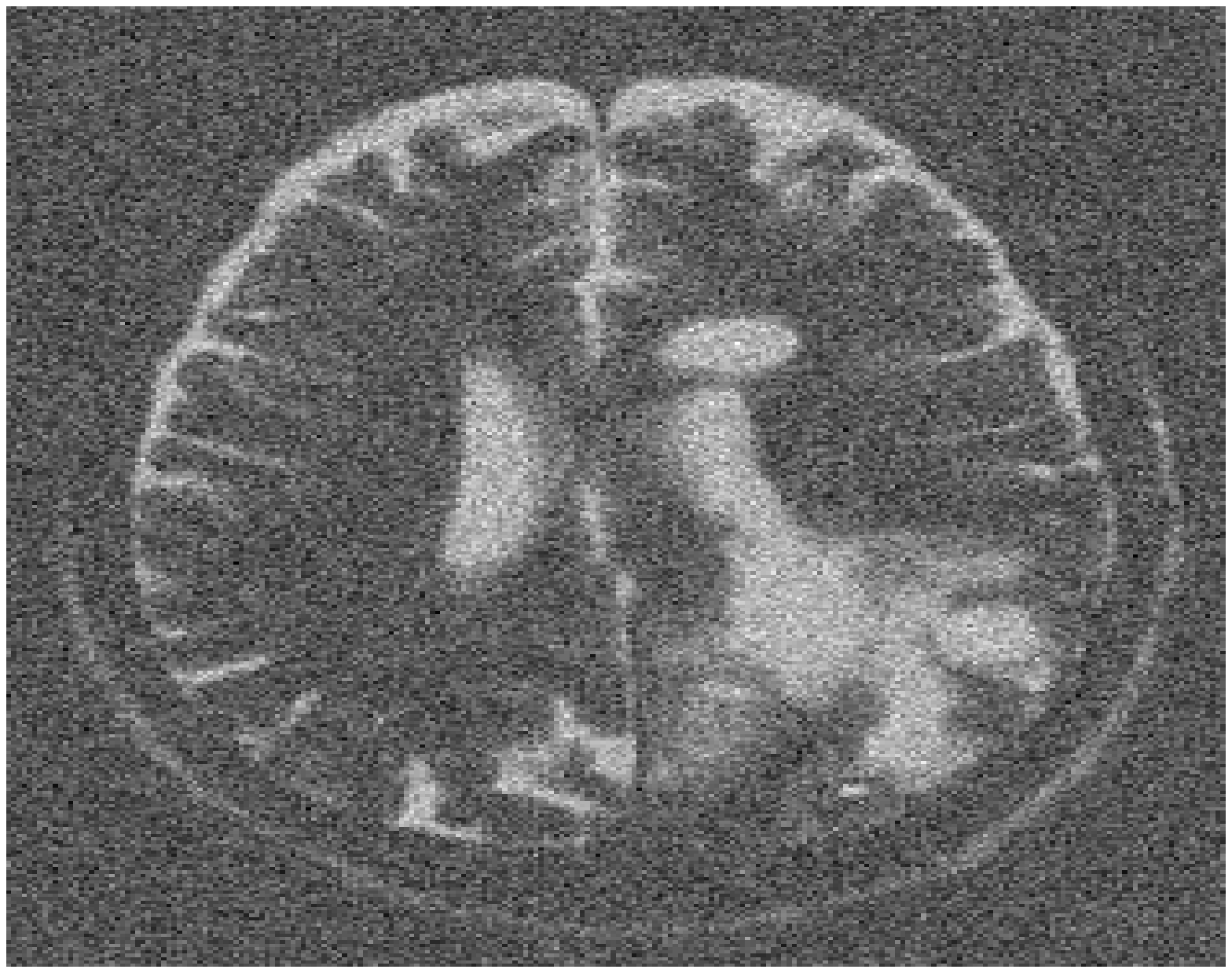} & 
	\includegraphics[width=22mm,height=22mm]{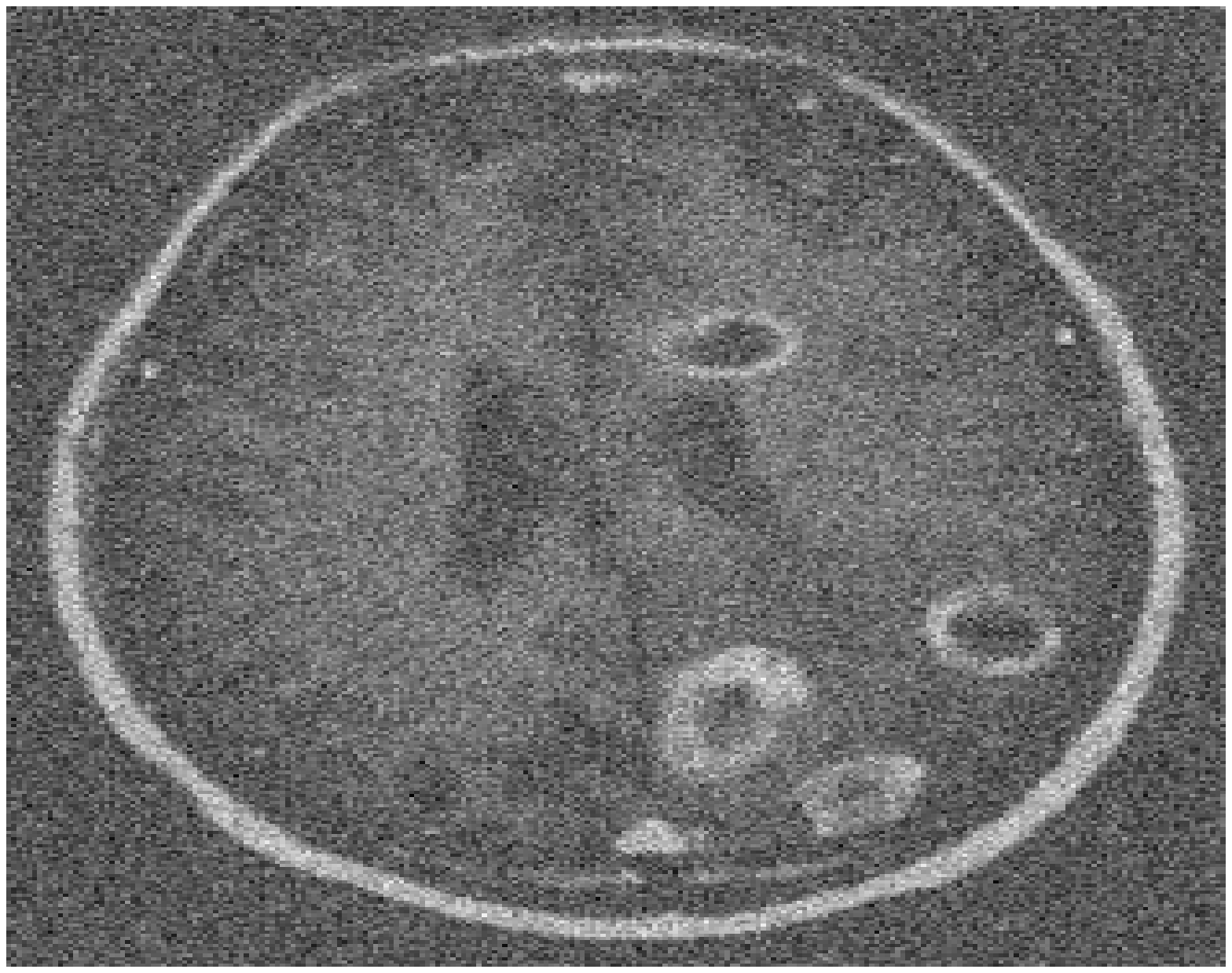} 
\\
 \multicolumn{3}{c}{(b)} 
\end{tabular}

\caption{Examples of images for data fusion and joint segmentation. a) PD (Proton Dennsity),T1 and T2-weighted slices of brain MR images. b) T1-weighted, T2-weighted and T1-weighted with contrast agent transversal slices of a 3D brain MR images with additive gaussian noise.}
\label{intro}
\end{figure}
\noindent
In this paper we consider the case where the measuring data systems can be assumed almost perfect and the observations are registered, which means that we can write :
\beq
g_i(r)=f_i(r)+\varepsilon_i(r), \qquad i=1,\dots,M
\eeq
\noindent
for $r\in \Rb^2$. If we have, for example, multispectral images, then $g_i$'s represent observations taken at different spectral bands. In a multimodal case, $g_i$ can represent CT and PET images. \\
\noindent 
Our aim is then to obtain a common segmentation of $N$ observations and to reconstruct $f_i,i=1,\dots,N$ at the same time. Figure \ref{intro} shows two examples of applications in MR images.\\ 
\noindent
This paper is organized as follows : In section 2 we introduce the common feature $z$, model the relation between the images $f_i$ to it through $p(f_i|z)$ and its proper characteristics through a prior law $p(z)$.
In section 3 we give detailed expressions of the \apost law and propose a general structure of the MCMC algorithm to estimate $\ufb$ and $\zb$. Finally, in section 4 we present some simulation results to show the performances of the proposed method.

\section{Modeling for Bayesian data fusion}

\noindent
In this paper we consider the model (2) where after discretization and using the notations $\gb_i=[g_i(1),\dots,g_i(S)]^T$ and $\ugb=(\gb_i)_{i=1,\dots,M}$, $\fb_i=[f_i(1),\dots,f_i(S)]^T$ and $\ufb=(\fb_i)_{i=1,\dots,M}$, and $\varepsilonb_i=[\varepsilon_i(1),\dots,\varepsilon_i(S)]^T$ and $\uvarepsilonb=(\varepsilonb_i)_{i=1,\dots,M}$, with $S$ the total number of pixels of the images $f_i$, we have :
\begin{eqnarray}
\ugb =  \ufb+\uvarepsilonb, \quad or \quad \gb_i  =  \fb_i+\varepsilonb_i, \qquad i=1,\dots,M 
\end{eqnarray}

\noindent
Within this model and assuming Gaussian independant noises,  $p(\varepsilonb_i)=\mathcal N(0,\sigma_{\varepsilon_i}^2 \Ib)$, we have

\beq
p(\ugb|\ufb)  = \prod_{i=1}^M p(\gb_i|\fb_i) = \prod_{i=1}^M p_{\varepsilon_i}(\gb_i-\fb_i)
\nonumber
\eeq

\noindent
As we want to reconstruct an image with statistically homogeneous regions, it is natural to introduce a hidden variable $\zb=(z(1),\dots,z(S)) \in \{1,\dots,K\}^S$ which represents a common classification of the images $\fb_i$. The problem is now to estimate the set of variables $(\ufb,\zb)$ using the Bayesian approach :

\beq
p(\ufb,\zb|\ugb) =  p(\ufb|\zb,\ugb)~p(\zb|\ugb) 
\eeq

\noindent
Thus to be able to give an expression for $p(\ufb,\zb|\ugb)$ using the Bayes formula, we need to define $p(\gb_i|\fb_i)$, $p(\fb_i|z)$, $p(\gb_i|z)$ and $p(\zb)$. \\
\noindent
Assuming $\varepsilonb_i$ centered, white and Gaussian, ans $S$ the number of pixels of an image, we have :
\beq
p(\gb_i|\fb_i) =  \mathcal N(\fb_i,\sigma_{\varepsilonb_i}^2 \Ib) 
=  \left( \frac{1}{2\pi \sigma_{\varepsilonb_i}^2} \right)^{\frac{S}{2}} \exp \left\{-\frac{1}{2\sigma_{\varepsilonb_i}^2}||\gb_i-\fb_i||^2 \right\}
\nonumber
\eeq
\noindent
To assign $p(\fb_i|\zb)$ we first define the sets of pixels which are in the same class :
\begin{eqnarray*}
R_k & = & \{r : z(r) = k\}, \qquad |R_k|=n_k \\
{\fb_i}_k & = & \{ f_i(r) : z(r)=k \}
\end{eqnarray*}

\noindent
Then we assume that all the pixels of an image $\fb_i$ which are in the same class will be characterized by a mean ${m_i}_k$  and a variance ${\sigma_i^2}_k$ :
\beq
p(f_i(r)|z(r)=k)=\mathcal N({m_i}_k,{\sigma_i^2}_k)
\nonumber
\eeq 

\noindent
With these notations we have :
\begin{eqnarray*}
p({\fb_i}_k) & = & \mathcal N({m_i}_k \bm{1},{\sigma_i^2}_k \Ib) \\
p(\fb_i |\zb) & = & \prod_{k=1}^K \Nc({m_i}_k \bm{1},{\sigma_i^2}_k \Ib) \\& = & 
\prod_{k=1}^K \left( \frac{1}{\sqrt{2\pi {\sigma_i^2}_k}}\right)^{n_k} \exp \left\{ - \frac{1}{2 {\sigma_i^2}_k} || {\fb_i}_k-{m_i}_k \bm{1} ||^2 \right\}, \qquad i=1,\dots,M.
\end{eqnarray*}

\noindent
The next step is to define $p(\gb_i|\zb)$. To do this we may use the relation (3) and the laws $p(\fb_i|\zb)$ and $p(\varepsilonb_i)$ to obtain
\beq
p(g_i(r)|z(r)=k)  =  \mathcal N({m_i}_k, {\sigma_i^2}_k + \sigma_{\varepsilon_i}^2) \\
\nonumber
\eeq

\noindent
Finally we have to assign $p(\zb)$. As we introduced the hidden variable $\zb$ for finding statistically homogeneous regions in images, it is natural to define a spatial dependency on these labels. The simplest model to account for this desired local spatial dependency is a Potts Markov Random Field model :
\beq
p(\zb) = \frac{1}{T(\alpha)} \exp \left\{ \alpha \sum_{r\in \mathcal S} \sum_{s \in \mathcal V(r)}\delta (z(r) - z(s)) \right\},
\nonumber
\eeq
where $\mathcal S$ is the set of pixels, $\delta(0)=1$, $\delta(t)=0$ if $t\neq0$, $\mathcal V(r)$ denotes the neighborhood of the pixel $r$ (here we consider a neighborhood of 4 pixels) and $\alpha$ represents the degree of the spatial dependency of the variable $\zb$. This parameter will be fixed for our algorithm.
\noindent
We have now all the necessary prior laws $p(\gb_i|\fb_i)$, $p(\fb_i|z)$, $p(\gb_i|z)$ and $p(\zb)$ and then we can give an expression for $p(\ufb,\zb|\ugb)$. However these probability laws have in general unknown parameters such as $\sigma^2_{\varepsilon_i}$ in $p(\gb_i|\fb_i)$ or ${m_i}_k$ and ${\sigma_i^2}_k$ in $p(\fb_i|\zb)$. In a full Bayesian approach, we have to assign prior laws to these "hyperparameters". Then the choice of prior laws for the hyperparameters is still an open problem. In \cite{Snoussi02c} the authors used differential geometry tools to construct particular priors which contain as particular case the entropic and conjugate priors. In this paper we choose this last one. \\
\noindent
Let $\mb_i=({m_i}_k)_{k=1,\dots,K}$ and $\sigmab_i^2=({\sigma_i^2}_k)_{k=1,\dots,K}$ be the means and the variances of the pixels in different regions of the images $\fb_i$ as defined before. We define $\thetab_i$ as the set of all the parameters which must be estimated :
\beq
\thetab_i = (\sigma_{\varepsilon_i}^2,\mb_i,\sigmab_i^2), \quad i=1,\dots,M
\nonumber
\eeq
and we note $\uthetab = (\thetab_i)_{i=1,\dots,M}$.
When applied the particular priors of (\cite{Snoussi02c}) for our case, we find the following conjugate priors :
\bit
\item Inverse Gamma $\mathcal{IG}(\alpha^{\varepsilon_i}_0,\beta^{\varepsilon_i}_0)$ and $\mathcal{IG}({\alpha_i}_0,{\beta_i}_0)$ respectively for the variances $\sigma_{\varepsilon_i}^2$ and ${\sigma^2_i}_k$, 
\item Gaussian $\mathcal N({m_i}_0,{\sigma^2_i}_0)$ for the means ${m_i}_k$. 
\eit
\noindent
The hyper-hyperparameters ${\alpha_i}_0$, ${\beta_i}_0$, ${m_i}_0$ and ${\sigma^2_i}_0$ are fixed and the results are not in general too sensitive to their exact values. However in case of noisy images we can constrain small value on ${\sigma^2_i}_0$ in order to force the reconstruction of homogeneous regions.


\section{A posteriori distributions for the Gibbs algorithm}
\noindent
The Bayesian approach consists now to estimate the whole set of variables $(\ufb,\zb,\uthetab)$ following the joint \apost distribution $p(\ufb,\zb,\uthetab | \ugb)$. It is difficult to simulate a joint sample $(\hat{\ufb},\hat{\zb},\hat{\uthetab})$ directly from his joint \apost distribution. However we can note that considering the prior laws defined before, we are able to simulate the conditionnal \apost laws $p(\ufb,\zb| \ugb,\uthetab )$ and $p(\uthetab | \ugb,\ufb,\zb)$. That is why we propose a Gibbs algorithm to estimate $(\hat{\ufb},\hat{\zb},\hat{\uthetab})$, splitting first this set of variables into two subsets, $(\ufb,\zb)$ and $(\uthetab)$ :
\beq 
p(\ufb,\zb | \ugb,\uthetab)=p(\ufb | \zb,\ugb,\uthetab)p(\zb | \ugb,\uthetab)
\nonumber
\eeq
\noindent
Then the sampling of this joint distribution is obtained by sampling first $p(\zb | \ugb,\uthetab)$ and then sampling $p(\ufb | \zb,\ugb,\uthetab)$.
We will now define the conditionnal \apost distribution we use for the Gibbs algorithm. \\

\noindent
{\bf Sampling $\zb|\ugb,\uthetab$ :}\\ 
for this step we have :
\beq
p(\zb | \ugb,\uthetab) \propto  p(\ugb |\zb,\uthetab)~p(\zb) 
 =  \prod_{i=1}^M p(\gb_i|\zb,\thetab_1) ~p(\zb)
\nonumber
\eeq
\noindent
As we choosed a Potts Markov Random Field model for the labels, we may note that an exact sampling of the \apost distribution $p(\zb | \ugb,\uthetab)$ is impossible. However we propose in section 4 a parallel implementation of the Gibbs sampling for resolving this problem.\\ \\ 
\noindent
{\bf Sampling $\fb_i|\gb_i,\zb,\thetab_i$ :}\\ 
We can write the \apost law $p(f_i(r)|g_i(r),z(r),\thetab_i)$ as follows :
\beq
p(f_i(r)|g_i(r),z(r)=k,\thetab_i) = \mathcal N({m_i}_k^{apost},{\sigma_i^2}_k^{apost})
\nonumber
\eeq
where
\beq
 {m_i}_k^{apost} = {\sigma_i^2}_k^{apost} \left( \frac{g_i(r)}{\sigma_{\varepsilon_i}^2} + \frac{{m_i}_k}{\sigma_i^2}_k \right)\quad and \quad {\sigma_i^2}_k^{apost}  =  \left( \frac{1}{\sigma_{\varepsilon_i}^2} + \frac{1}{\sigma_i^2}_k\right)^{-1}
\nonumber
\eeq

\noindent
{\bf sampling $\thetab_i|\fb_i,\gb_i,\zb$ :} \\ 
We have the following relation :
\beq 
p(\thetab_i|\fb_i,\gb_i,\zb) \propto p({\sigma_{\varepsilon_i}^2} | \fb_i,\gb_i)\hbox{ }p(\mb_i,\sigmab^2_i | \fb_i,\zb)
\nonumber
\eeq
\noindent 
and using again the Bayes formula, the \apost distributions are calculated from the prior selection fixed before and we have \\
\indent
- ${m_i}_k|\fb_i,\zb,{\sigma_i^2}_k,{m_i}_0, {\sigma^2_i}_0 \sim \mathcal N({\mu_i}_k,{v_i^2}_k)$, with
\beq
{\mu_i}_k  =   {v_i^2}_k \left( \frac{{m_i}_0}{{\sigma^2_i}_0} + \frac{1}{{\sigma^2_i}_k} \sum_{r \in R_k} f_i(r)   \right) \quad and \quad
{v_i^2}_k  =  \left( \frac{n_k}{{\sigma_i^2}_k} + \frac{1}{{\sigma^2_i}_0}\right)^{-1}
\nonumber
\eeq 

\indent 
- ${\sigma_i^2}_k |\fb_i,\zb,{\alpha_i}_0,{\beta_i}_0 \sim \mathcal{IG}({\alpha_i}_k,{\beta_i}_k)$, with
\beq
{\alpha_i}_k  = {\alpha_i}_0 + \frac{n_k}{2} \quad and \quad
{\beta_i}_k  =  {\beta_i}_0 + \frac{1}{2}\sum_{r \in R_k} (f_i(r)-{m_i}_k)^2
\nonumber
\eeq 
\indent
- $\sigma_{\varepsilon_i}^2 |\fb_i,\gb_i \sim \mathcal{IG}(\nu_i,\Sigma_i)$, with
\beq
\nu_i  =  \frac{S}{2} + \alpha^{\varepsilon_i}_0, \qquad S=\mbox{ number of pixels} \quad and \quad
\Sigma_i  =  \frac{1}{2} ||\gb_i-\fb_i||^2 + \beta^{\varepsilon_i}_0
\eeq


\section{Parallel implementation of the Gibbs algorithm}
As we choosed a first order neighborhood system for the labels, we may also note that it is possible to implement the Gibbs algorithm in parallel. Indeed, we can decompose the whole set of pixels into two subsets forming a chessboard (see figure 2). In this case if we fix the black (respectively white) labels, then the white (respectively black) labels become independant. 
\begin{figure}[h]
\centering
\includegraphics[width=40mm,height=20mm]{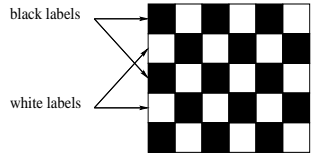}
\caption{Chessboard decomposition of the labels $z$}
\end{figure}
This decomposition reduces the complexity of the Gibbs algorithm because we can simulate the whole set of labels in only two steps. The Parallel Gibbs algorithm we implemented is then the following : given an initial state $(\hat{\thetab_1},\hat{\thetab_2},\hat{\zb})^{(0)}$,

\begin{center}
{\bf  Parallel Gibbs sampling} \\
\begin{tabular}{l}
\hline repeat until convergence \\
 \begin{tabular}{l} 
 		1. simulate $\hat{\zb_B}^{(n)} \sim p(\zb|\hat{\zb_W}^{(n-1)},\ugb,\hat{\uthetab}^{(n-1)})$ \\
		$\quad$ simulate $\hat{\zb_W}^{(n)} \sim p(\zb|\hat{\zb_B}^{(n)},\ugb,\hat{\uthetab}^{(n-1)})$ \\
 		$\quad$ simulate $\hat{\fb_i}^{(n)} \sim p(\fb_i | \gb_i,\hat{\zb}^{(n)},\hat{\thetab_i}^{(n-1)})$ 
	\end{tabular} \\
 \begin{tabular}{l}
  2. simulate $\hat{\thetab_i}^{(n)} \sim p(\thetab_i|\hat{\fb_i}^{(n)},\hat{\zb}^{(n)},\gb_i)$ 
 	\end{tabular} \\
\hline
\end{tabular}
\end{center} 


\section{Simulation and results}

\noindent
Here we illustrate two examples of MRI images : PD, T1-weightd and T2-weighted slices of a MR brain image, which are ($188\times193$) images for the first example, and T1-weighted,T2-weightedand T1-weighted with contrast agent slices of a MR brain image, which are ($289\times236$) images for the second example. In this last we have added a gaussian noise.


\begin{figure}[h]
\centering
\begin{minipage}[t]{3cm}
\begin{tabular}{c}
	\begin{tabular}{ccc}
		\includegraphics[width=22mm,height=22mm]{ftt3} & 
		\includegraphics[width=22mm,height=22mm]{ftt1} & 
		\includegraphics[width=22mm,height=22mm]{ftt2}  
		\\
		\includegraphics[width=22mm,height=22mm]{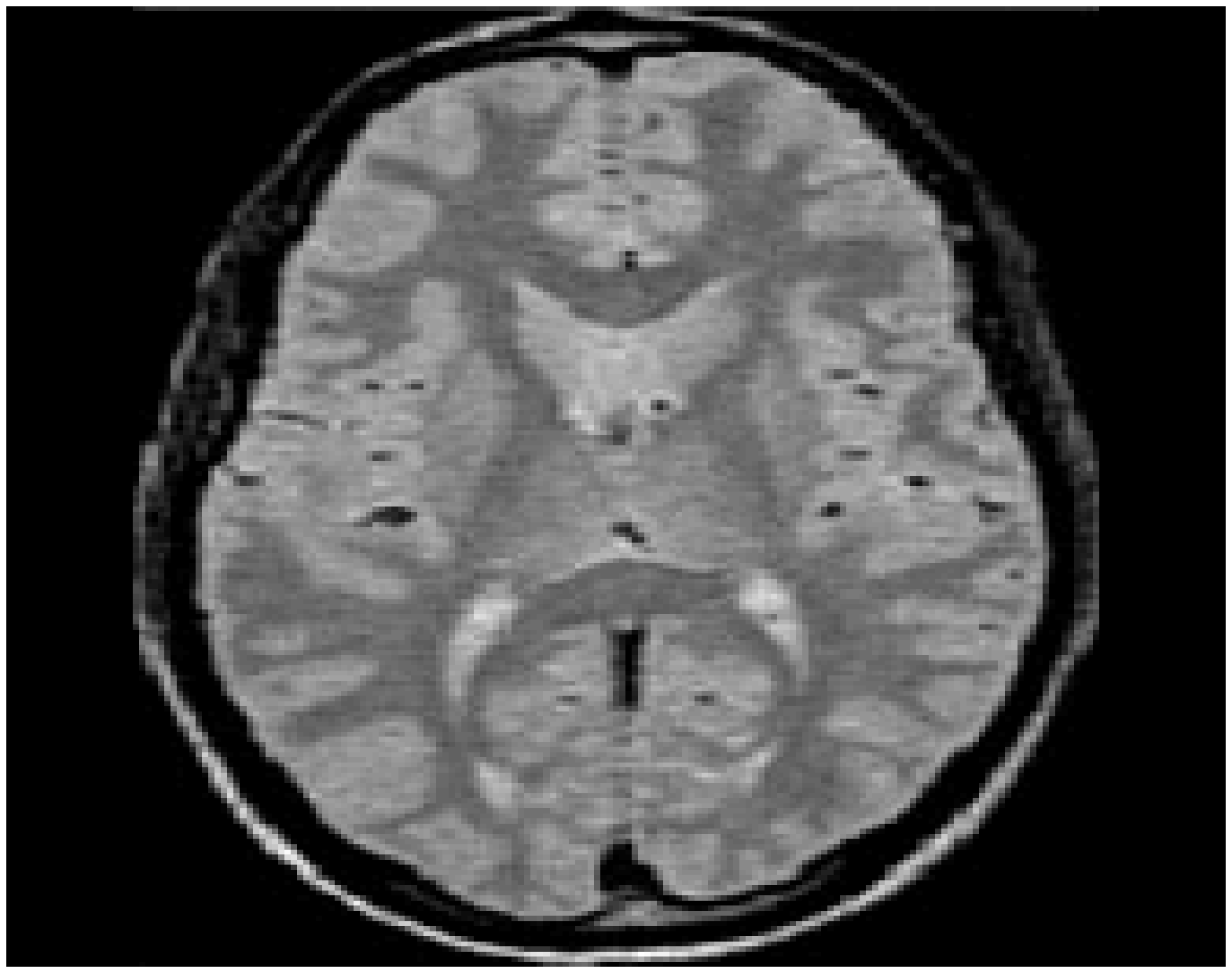} & 
		\includegraphics[width=22mm,height=22mm]{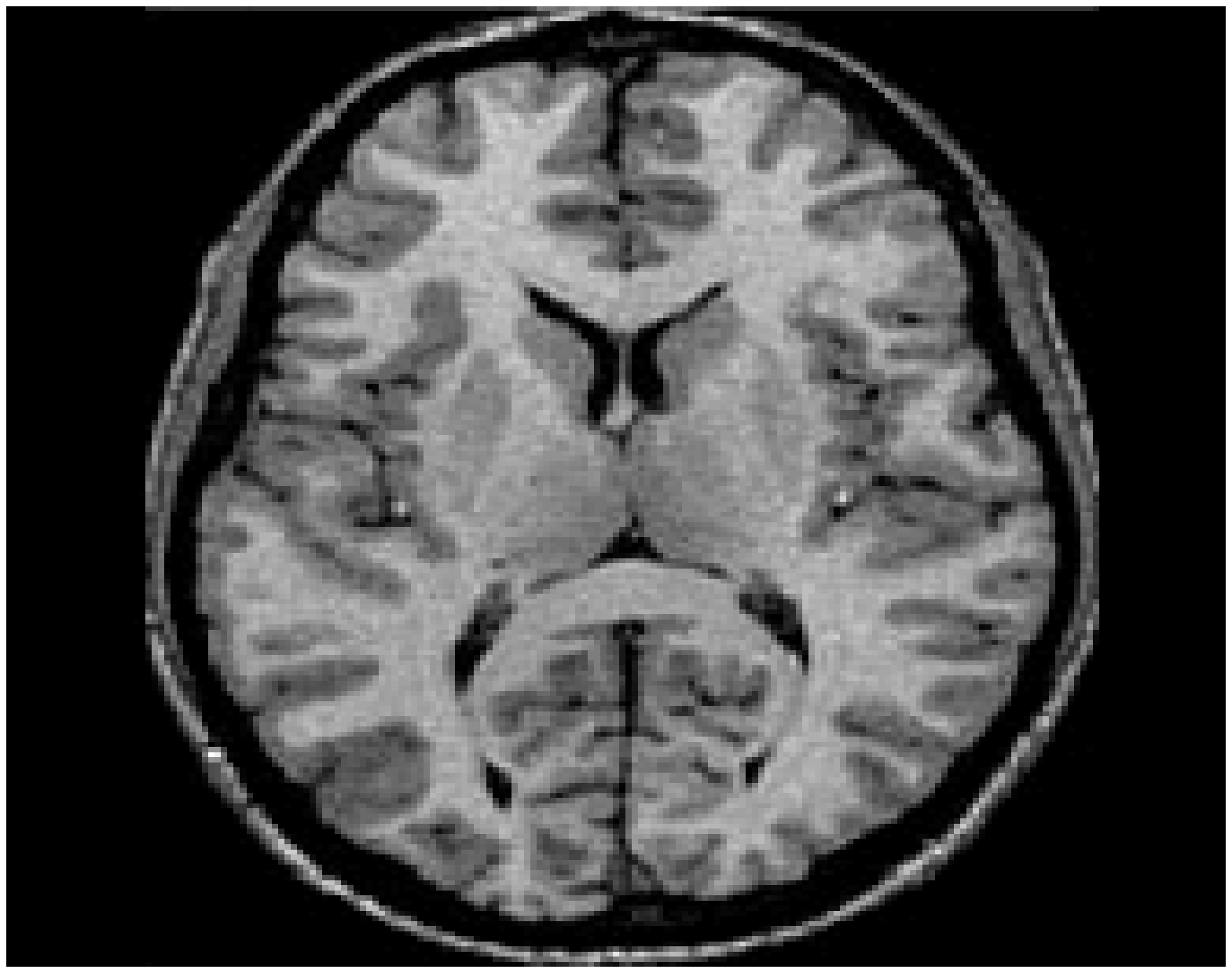} & 
		\includegraphics[width=22mm,height=22mm]{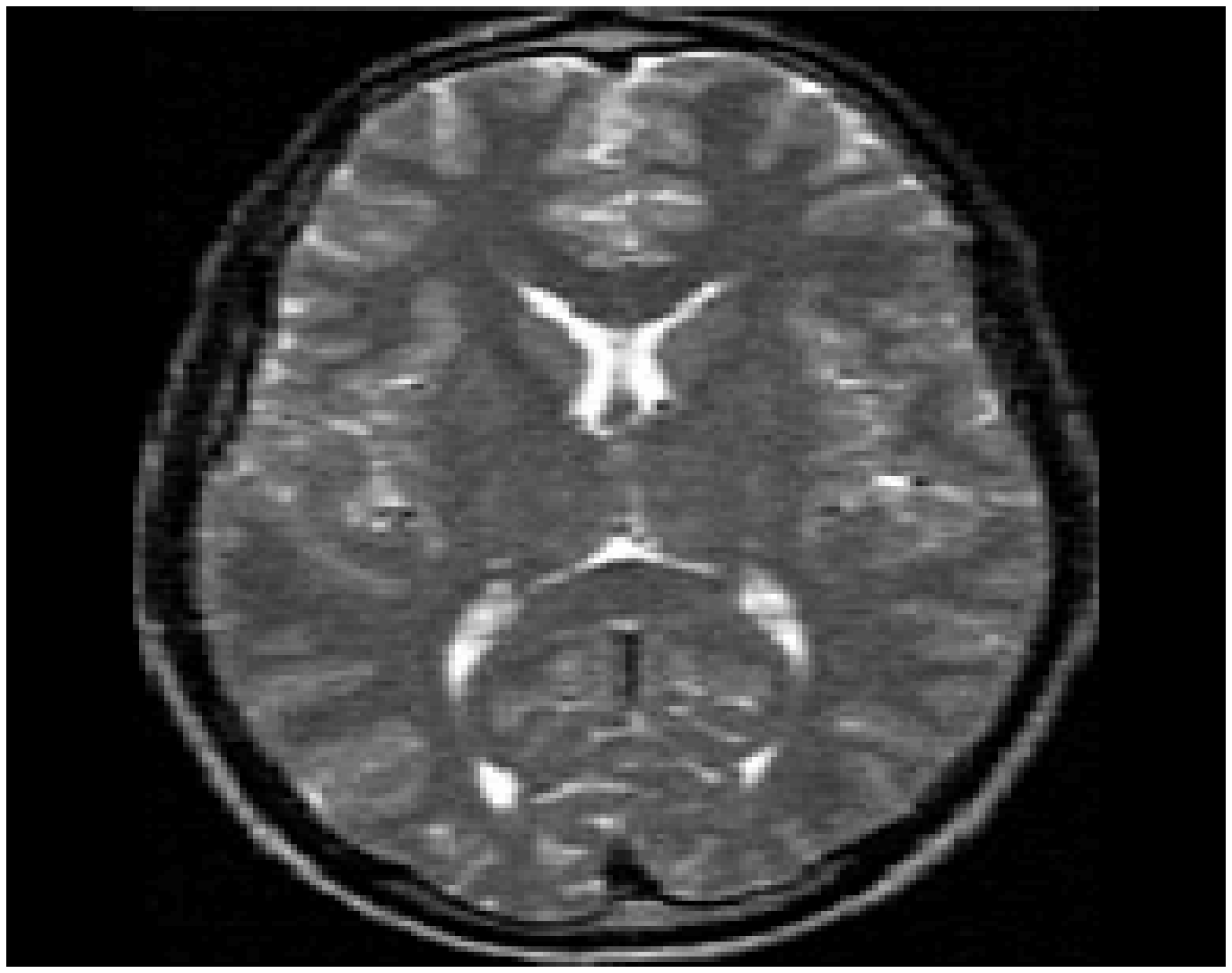} 
	\end{tabular}
	\\
	(a)
\end{tabular}
\end{minipage}
\hspace{38mm}
\begin{minipage}[t]{5cm}
	\begin{tabular}{c}
		\includegraphics[width=45mm,height=45mm,]{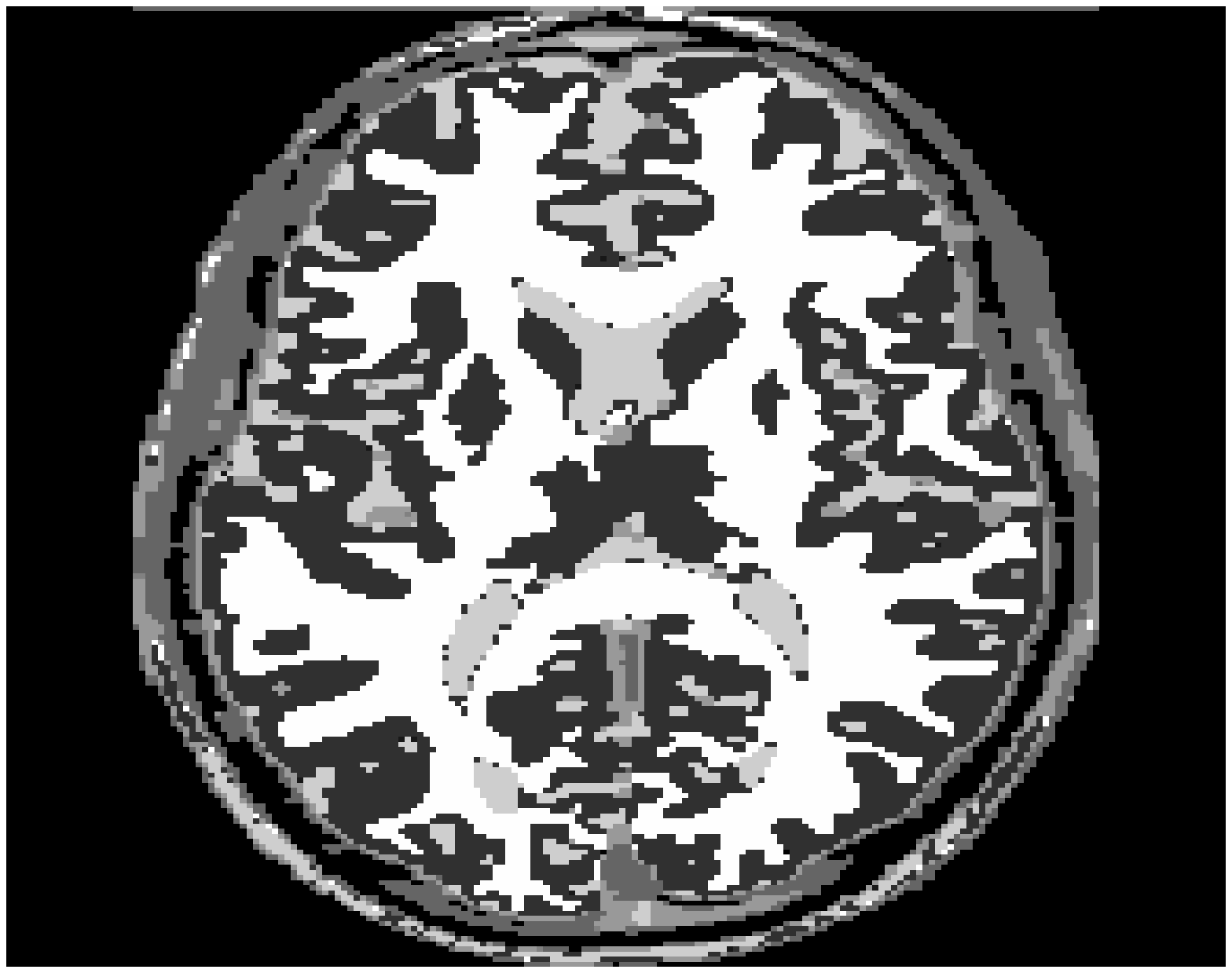} \\
		(b)
	\end{tabular}
	
	\end{minipage}
	
\caption{Results of data fusion from MRI images. a)observations (up) and reconstructed images (down) . b) result of data fusion with 7 labels.}
\label{ftt}
\end{figure}

\begin{figure}[h]
\centering
\begin{minipage}[t]{3cm}
\begin{tabular}{c}
	\begin{tabular}{ccc}
		\includegraphics[width=22mm,height=22mm]{tumorbruite1} & 
		\includegraphics[width=22mm,height=22mm]{tumorbruite2} & 
		\includegraphics[width=22mm,height=22mm]{tumorbruite3}  
		\\
		\includegraphics[width=22mm,height=22mm]{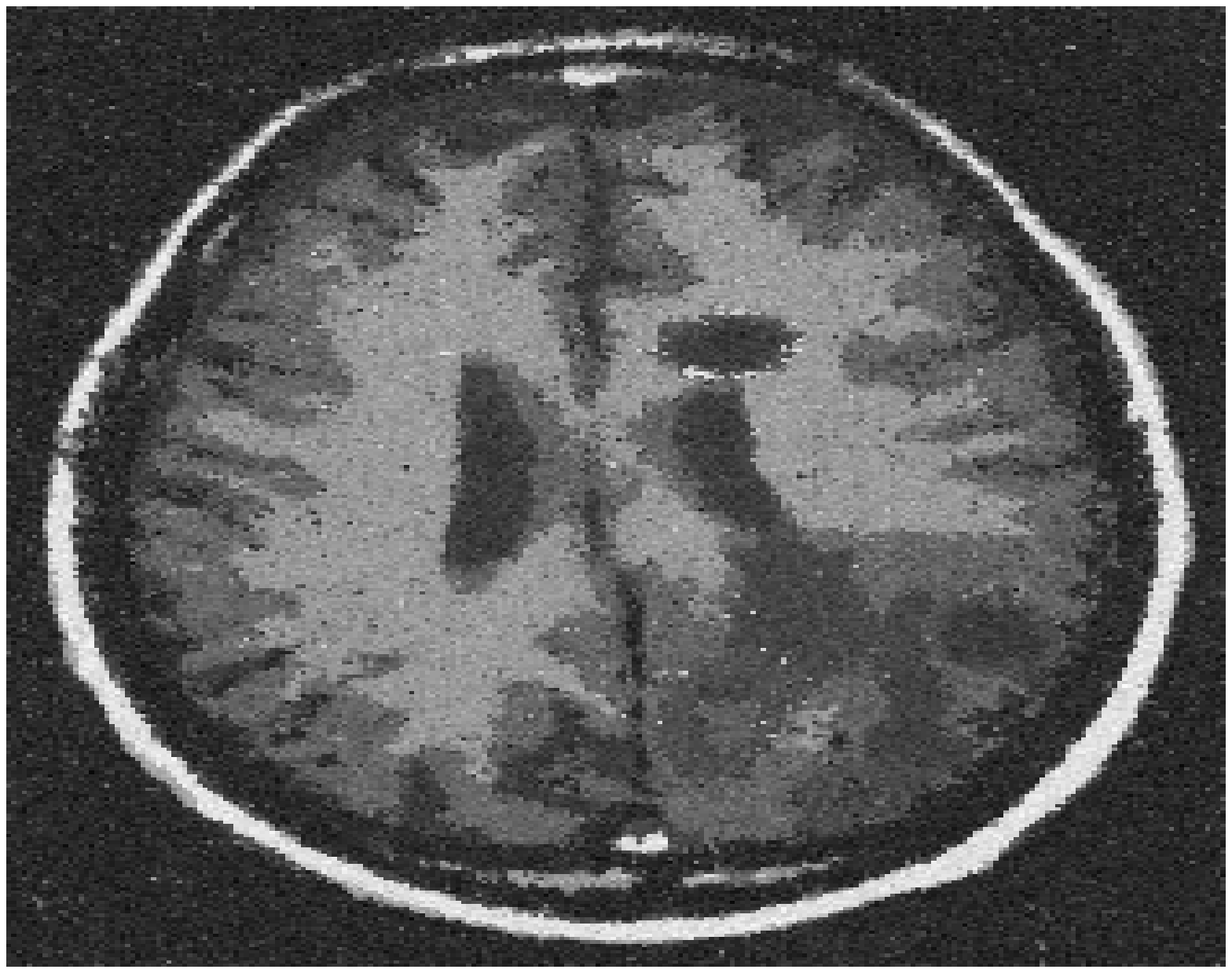} & 
		\includegraphics[width=22mm,height=22mm]{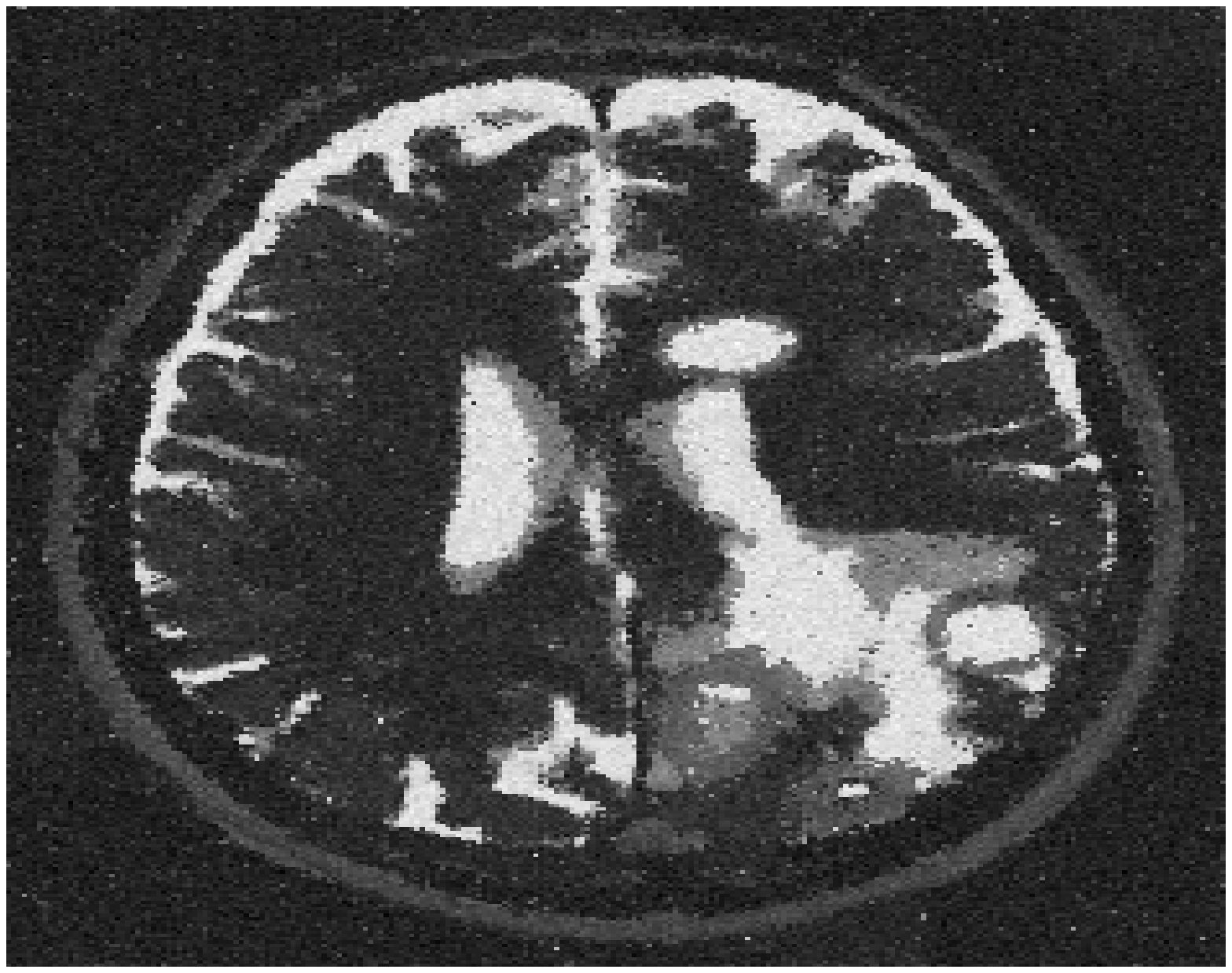} & 
		\includegraphics[width=22mm,height=22mm]{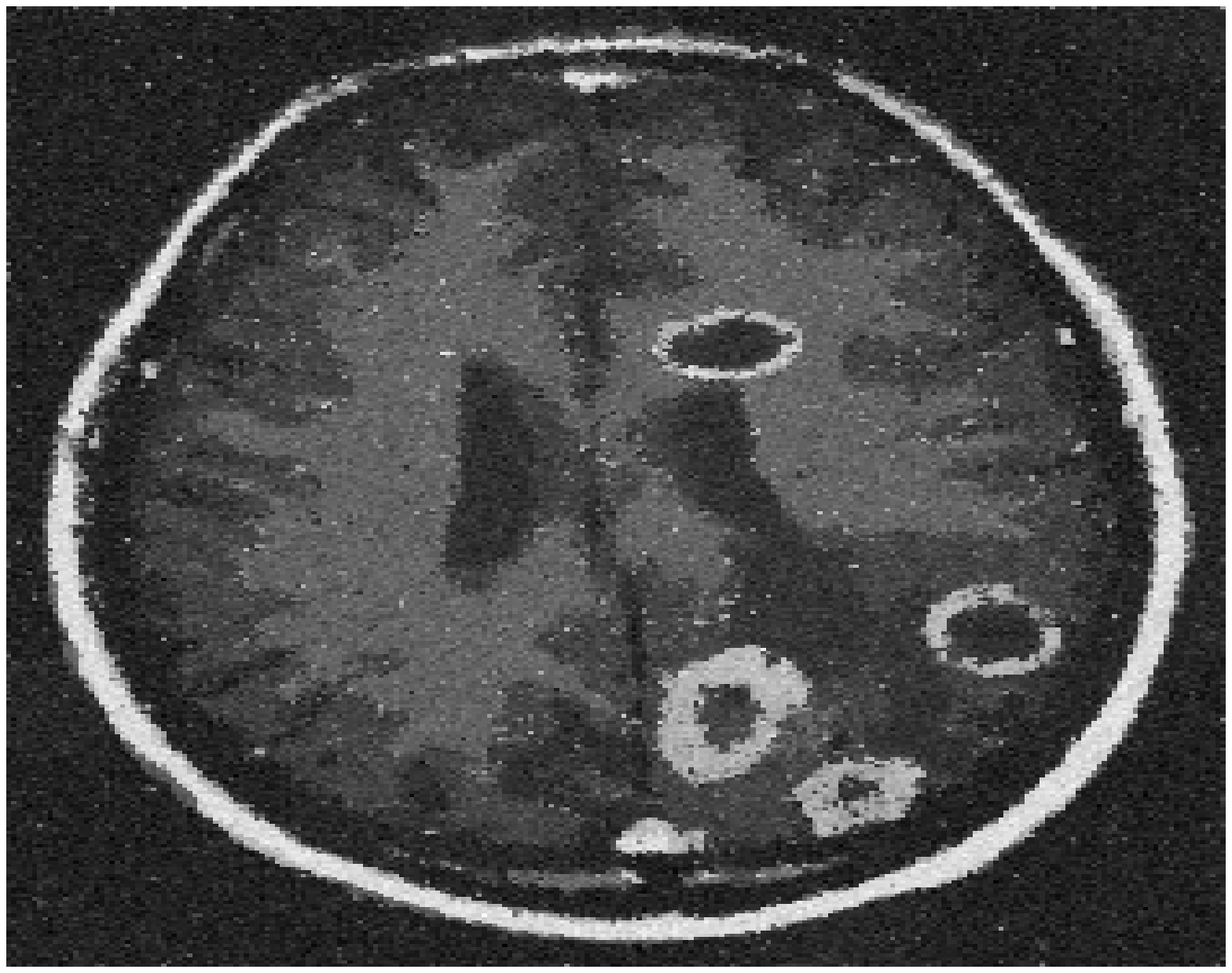} 
	\end{tabular}
	\\
	(a)
\end{tabular}
\end{minipage}
\hspace{38mm}
\begin{minipage}[t]{5cm}
	\begin{tabular}{c}
		\includegraphics[width=45mm,height=45mm,]{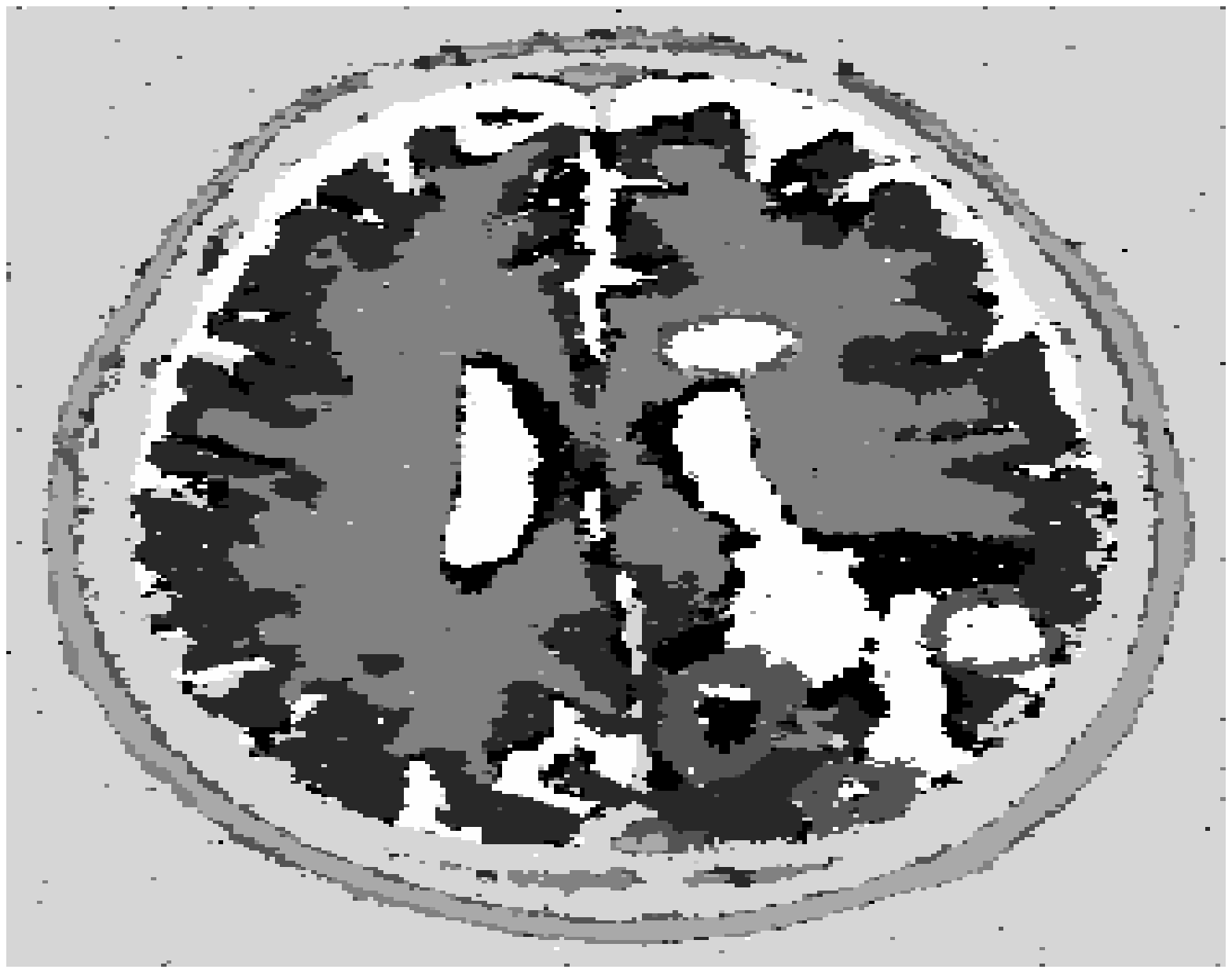} \\
		(b)
	\end{tabular}
	
	\end{minipage}
	
\caption{Results of data fusion from MRI images. a)observations (up) and reconstructed images (down) . b) result of data fusion with 7 labels.}
\label{tumor}
\end{figure}

\noindent
Figures \ref{ftt} and \ref{tumor} show the data fusion result of the proposed method. As it is seen on these figures the fusionned segmentations we obtain contain all the regions and boundaries of the observations, but we have not yet compared with other methods to see the performances of our algorithm. Also the presence of noise in figure \ref{tumor} do not really affect the result of segmentation and, at the same time, the proposed algorithm give visibly improved reconstructed images.
\noindent

%
\noindent
In both applications we have satisfactory results of image fusion, even when images present a great number of homogeneous regions and boundaries. Note also that in both applications we fixed a prior small value of ${\sigma^2_i}_0$ to improve the reconstructed images. Another way may be the introduction of some local spatial dependency between the neighboring pixels of images $f_i(r)$. This point is under development and we will report soon on the results.

\section{Conclusion}
\noindent
We proposed a Bayesian method for data fusion of images, with a Potts Markov Random Field model on the hidden variable $\zb$. We illustrated how a joint segmentation and reconstruction can be obtained in case of MRI images. 
We showed then how reconstruction and fusion can be computed at the same time using a MCMC algorithm. We considered the case of noisy images and showed that the joint segmentation is not greatly affected. This method give an unsupervised segmentation which do not take into account particular shapes and then can give good results in many different cases. However we assume for the moment that the observed images are registered. We think that this modelisation is promising for introducing registration operators $H_i$ and then implementing common segmentation and registration at the same time. Another perspective is to introduce spatial dependency directly on the images $\fb_i$ for involving the reconstruction.

\nocite{Held97,Gindi93,Gautier98,Hebert89,Robert96}

\small 

\bibliographystyle{ieeetr}

\bibliography{revuedef,biben,baseAJ,baseKZ,gpipubli,amd99,miccai04}

\edoc

%% file: Input/abrmath.tex
\def\th{{\mathrm{th}}}                

\newsavebox{\fminibox}
\newlength{\fminilength}
\newenvironment{fminipage}[1][\linewidth]
  {\setlength{\fminilength}{#1}
   \begin{lrbox}{\fminibox}\begin{minipage}{\fminilength}}
  {\end{minipage}\end{lrbox}\noindent\fbox{\usebox{\fminibox}}}

  \def\+{^\dagger}

\def\nequiv{\not\kern-.05em\equiv}
\def\egal{\kern-.5em=\kern-.5em}        
\def\propt{\kern-.2em\propto\kern-.2em} 

\def\argmax{\mathop{\mathrm{arg\,max}}} 
\def\argmin{\mathop{\mathrm{arg\,min}}} 
\def\intdouble{\int\kern-0.3em\int}
\def\inttriple{\int\kern-0.3em\int\kern-0.3em\int}

\def\rond#1{\overset{\kern-0.33em~_\circ}{#1}}
\def\rondit[#1]#2{\overset{\kern#1~_\circ}{#2}}


%% file: Input/beginend.tex
\def\babs{\begin{abstract}}             \def\eabs{\end{abstract}}
\def\barr{\begin{array}}                \def\earr{\end{array}}
\def\bcc{\begin{center}}                \def\ecc{\end{center}}

\def\bdes{\begin{description}}          \def\edes{\end{description}}
\def\bdoc{\begin{document}}             \def\edoc{\end{document}}
\def\ben{\begin{enumerate}}             \def\een{\end{enumerate}}
\def\beqn{\begin{eqnarray}}             \def\eeqn{\end{eqnarray}}
\def\beqnl#1{\beqn\label{#1}}           \def\eeqnl#1{\label{#1}\eeqn}
\def\beqnx{\begin{eqnarray*}}           \def\eeqnx{\end{eqnarray*}}
\def\bseqn{\begin{subeqnarray}}         \def\eseqn{\end{subeqnarray}}
\def\beq#1\eeq{\begin{equation}#1\end{equation}}
\def\bal#1\eal{\begin{align}#1\end{align}}
\def\balx#1\ealx{\begin{align*}#1\end{align*}}
\def\beqx{$$}                           \def\eeqx{$$}
\def\bfig{\protect\begin{figure}}       \def\efig{\protect\end{figure}}
\def\bfigx{\protect\begin{figure*}}     \def\efigx{\protect\end{figure*}}
\def\bfigt{\protect\begin{figurette}}   \def\efigt{\protect\end{figurette}}
\def\bfl{\begin{flushleft}}             \def\efl{\end{flushleft}}
\def\bfr{\begin{flushright}}            \def\efr{\end{flushright}}
\def\bit{\begin{itemize}}               \def\eit{\end{itemize}}
\def\bmi{\begin{minipage}}              \def\emi{\end{minipage}}
\def\bfmi{\begin{fminipage}}            \def\efmi{\end{fminipage}}
\def\bpic{\begin{picture}}              \def\epic{\end{picture}}
\def\bqu{\begin{quote}}                 \def\equ{\end{quote}}
\def\bqun{\begin{quotation}}            \def\equn{\end{quotation}}
\def\bsl{\begin{slide}}                 \def\esl{\end{slide}}
\def\btabb{\begin{tabbing}}             \def\etabb{\end{tabbing}}
\def\btabl{\begin{table}}               \def\etabl{\end{table}}
\def\btablx{\begin{table*}}             \def\etablx{\end{table*}}
\def\btab{\begin{tabular}} 
\def\btabu{\begin{tabular}}             \def\etabu{\end{tabular}}
\def\btabx{\begin{tabular*}}            \def\etabx{\end{tabular*}}
\def\bbib{\begin{thebibliography}{999}} \def\ebib{\end{thebibliography}}
\def\bver{\begin{verbatim}}             \def\ever{\end{verbatim}}
\def\bca{\begin{cases}}                  \def\eca{\end{cases}}